# The sensitivity and behaviour of the curvature in the échelle diagram of red-giant stars

S. Hekker[1,2], Y. Elsworth[3], S. Basu[4], F. Ahlborn[1], W. H. Ball[3,5], E. P. Bellinger[4], L. Buchele[6,1,2], and F. Espinoza-Rojas[1,2]

[1] Heidelberger Institut für Theoretische Studien, Schloss-Wolfsbrunnenweg 35, 69118 Heidelberg, Germany
   e-mail: `saskia.hekker@h-its.org`
[2] Center for Astronomy (ZAH/LSW), Heidelberg University, Königstuhl 12, 69117 Heidelberg, Germany
[3] School of Physics and Astronomy, University of Birmingham, Birmingham B5 2TT, UK
[4] Department of Astronomy, Yale University, PO Box 208101, New Haven, CT 06520-8101, USA
[5] Advanced Research Computing, University of Birmingham, Birmingham B5 2TT, UK
[6] Institute of Science and Technology Austria (ISTA), Am Campus 1, 3400 Klosterneuburg, Austria

**ABSTRACT**

*Context.* In the convective envelopes of relatively cool (surface temperature ≲ 6700 K) stars, oscillations are excited by turbulent convection. In these so-called solar-like oscillators, radial oscillation modes appear at nearly equally spaced frequencies. This spacing is referred to as the 'large frequency separation'. Deviations from equally-spaced frequencies are a result of the internal structure of a star being different from a sphere of ideal gas at constant temperature. Hence, these deviations provide information on the internal structure of the star.
*Aims.* In this work, we investigate the second-order (quadratic) deviation from uniform spacing, referred to as curvature. We aim to provide homogeneous values for observed red-giant stars, understand differences between the results from observations and predictions from stellar models, and reveal the connection between curvature and stellar structure.
*Methods.* We used *Kepler* data of red-giant stars and computed the curvature for several thousand stars. We compared these to the curvature derived from MESA models. We subsequently investigated the trends and differences between results from observations and models. Finally, we computed sensitivity kernels to identify the stellar region(s) to which the curvature is most sensitive and performed a glitch analysis.
*Results.* We found that the curvature is sensitive to evolutionary phase and mass. Interestingly, the observed values and values from models show some discrepancies. Including the surface effect in the model frequencies reduces the discrepancies, though introduces a frequency-dependent over- or under-estimation of the curvature from the models compared to the observations. From the kernels, we confirmed that the curvature is mostly sensitive to the near-surface layers of the star. The glitch analysis shows that in theory this provides information on the location and strength of the He I and H I ionisation layers.
*Conclusions.* The curvature provides a probe into the near-surface structure of the star. The deviations between the curvature derived from observations and models call henceforth for improvements in the near-surface layers of stellar models.

## 1. Introduction

The *Kepler* mission observed many stars, which enabled the detection and interpretation of solar-like oscillations in red giants. Solar-like oscillations are stochastically excited by the turbulent convection in the outer layers of the stars. The restoring force of the radial oscillation ($\ell = 0$) modes is the pressure gradient. These oscillations are referred to as pressure (p) modes. In the case of non-radial modes ($\ell > 0$) the pressure modes in the outer parts of the star couple with modes in the core of the red-giant stars where buoyancy is the restoring force. These oscillations are referred to as mixed modes. The set of radial and non-radial modes are thus sensitive to different parts of the stars, which provides a wealth of information on the stellar structure, e.g. Bedding et al. (2011); Beck et al. (2012); Stello et al. (2016); Mosser et al. (2017); Hekker et al. (2018); Pinçon et al. (2020); Lindsay et al. (2023).

An essential ingredient for the use of oscillations in determining the internal stellar structure is the identification of radial order ($n$), spherical degree ($\ell$) and for non-radial modes the azimuthal order ($m$). In the case of solar-like oscillations in red-giant stars, the frequencies of low-degree ($\ell = 0, 1, 2, 3$) modes appear in a pattern that shows similarities — though with some changes due to the appearance of mixed modes — with the low-degree mode pattern observed in the Sun. This pattern can be described by the asymptotic relation as developed by Tassoul (1980), as the observed low-degree modes are in the asymptotic limit with $\ell \ll n$. However, the radial order of the observed modes decreases with evolution, and hence for red giant stars it becomes questionable if the asymptotic approximation is still valid. In practice, this means that increasing deviations from the asymptotic relation are expected with the evolution of stars on the red-giant branch (RGB). These deviations reduce again in the core-helium burning phase as the observed oscillations in these stars occur at higher radial orders compared to the oscillations in stars high on the RGB.

To account for the expected deviations from the asymptotics, Mosser et al. (2011) adapted the asymptotic relation developed

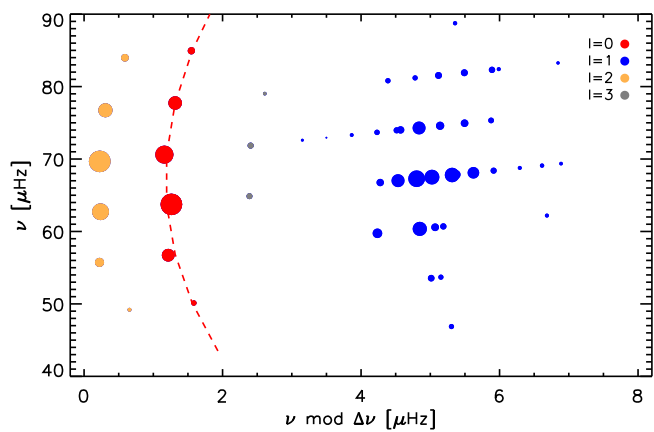

**Fig. 1.** Échelle diagram of KIC 2443903 with the degrees of the modes indicated with different colours as per the legend. The curvature, i.e. not straight vertical orientation of the radial modes is highlighted with the red dashed line.

by Tassoul (1980) by adding a second order term (indicated in blue):

$$\nu_{n,\ell} = \Delta\nu\left(n + \frac{\ell}{2} + \epsilon_p + \frac{\alpha}{2}(n - n_{max})^2\right) + d_{0,\ell}, \quad (1)$$

for the eigenfrequencies $\nu_{n,\ell}$, where $\Delta\nu$ is the asymptotic large frequency separation, i.e. the typical frequency difference between modes of the same degree and consecutive radial order, $\epsilon_p$ is a phase term and $d_{0,\ell}$ a small correction to the leading order asymptotics, which is zero for $\ell = 0$ modes. In the second-order term, indicated in blue, $\alpha$ accounts for the frequency-dependent deviation from the regular pattern set by $\Delta\nu$ and is often referred to as '*curvature*'. Finally, $n_{max}$ is the radial order[1] of the frequency of maximum oscillation power ($\nu_{max}$). See Fig. 1 for a so-called échelle diagram in which the curvature is illustrated.

Mosser et al. (2011, 2012, 2013b,a) have derived several prescriptions for $\alpha$ in stars with $\Delta\nu < 20\ \mu$Hz. The most recent one is:

$$\alpha = 0.015\Delta\nu^{-0.32} = 0.09 n_{max}^{-0.96} \approx 2 a_{RG} n_{max}^{-1}, \quad (2)$$

where $\Delta\nu \propto \nu_{max}^{0.75}$ (e.g. Hekker et al. 2011) and the value of $a_{RG} = 0.038 \pm 0.002$ is obtained from a fit, see Mosser et al. (2013b,a) for more details, and Vrard et al. (2015) for an application.

In this work, we aim to take the next step in the determination and interpretation of the curvature. Our first aim is to determine $\alpha$ and $\Delta\nu$ in a homogeneous manner for all red-giants that exhibit a sufficient number of detected radial modes (see Section 2 for a more precise discussion of the criteria). We then compare the curvature values obtained from observed red giants with curvature values obtained from models (Sections 3 & 4). Finally, we compute the sensitivity of the curvature to the internal structure of the stars (Section 5) and link this to features of the internal structure (Section 6). We finish with our discussion and conclusions in Section 7.

## 2. Analysis of curvature in observations

For the observations, we used the *Kepler* lightcurves prepared with the method described by Handberg & Lund (2014) for the 6661 stars from the APOKASC DR2 catalogue (Pinsonneault et al. 2018). To be sure that we compare like-for-like, we used only stars for which the available timeseries is at least 1200 days long with a filling factor larger than 65% throughout this paper. These timeseries data were subsequently analysed using the TACO (Tools for the Automated Characterisation of Oscillations) code (Themeßl et al. 2020, Hekker et al. in prep). In TACO, the timeseries data are converted to the frequency domain using a Fourier transformation. TACO then performs a global fit to the resulting power density spectrum (PDS) following the prescription in Kallinger et al. (2014). This global fit includes three Lorentzian-like functions each with an exponent of four to account for background signal (instrumentation and granulation at different scales), white noise and a Gaussian-shaped oscillation power excess. The power density spectrum is subsequently normalized by the background + white noise part of this global fit. This leaves us with a normalized power density spectrum, where the oscillations are the dominant signal. Finally, TACO fits all significant peaks in the relevant frequency range of the PDS and locates the radial modes in this normalized spectrum using the regularity in the frequencies of the oscillation modes as expressed in Eq. (1).

---
[1] The value of $n_{max}$ is not necessarily an integer.

For a meaningful comparison of the values of $\alpha$ between different stars as well as between stars and models, we compiled a sample of stars that have at least six radial modes observed and identified. For stars with more than six radial modes, we used the six modes closest to $\nu_{max}$. This provides us with a homogeneous sample. The selection of six modes is empirically determined and provides a balance between having enough stars with enough radial modes identified and having robust determinations of $\alpha$, i.e. a sufficient number of modes to measure the curvature.

Based on these requirements and omitting some stars with unusual features, such as high complexity (Choi et al. 2025) in their power density spectra, we present here a sample of 2242 red-giant-branch (RGB) stars and 1698 core-helium-burning (CHeB) stars. The evolutionary stages are taken from the APOKASC DR2 catalogue (Elsworth et al. 2019).

### 2.1. Determination of $\alpha$

The second-order term in Eq. (1) introduces two additional parameters ($\alpha$ & $n_{max}$). To avoid explicit correlations with other parameters such as $\nu_{max}$, we opted here to eliminate $n_{max}$ from the fitting. We therefore rewrote Eq. (1) for radial modes, i.e. $\ell = 0$, as a general second-order equation:

$$\nu_{n,\ell=0} = An^2 + Bn + C, \quad (3)$$

with

$$A = \Delta\nu\frac{\alpha}{2} \quad (4)$$

$$B = \Delta\nu(1 - \alpha n_{max}) \quad (5)$$

$$C = \Delta\nu\left(\epsilon_p + \frac{\alpha n_{max}^2}{2}\right) \quad (6)$$

To obtain $\alpha$, we fit the radial mode frequencies using Eq. (3) to obtain values for $A$, $B$, and $C$. To use $A$ to derive an estimate of $\alpha$ we need a value for $\Delta\nu$, which we determined from a linear fit of the frequencies to the radial orders ($\nu_{n,\ell=0} = \Delta\nu(n + \epsilon_p)$). In essence, this minimises the variance in $\nu$ modulo $\Delta\nu$, i.e., the ridges in an échelle diagram (see Fig. 1) are assumed to be vertical straight lines to first order. We note here that this is a different way of fitting the Universal pattern (Eq. (1)) as compared to earlier works by e.g. Mosser et al. (2013b), which may alter the values obtained for $\alpha$ as well as $\Delta\nu$. The approach we adopted here does not require a value of $n_{max}$, though it implicitly leads to the curvature to be somewhere around the mean value of $n$. This implicit value of $n_{max}$ only depends on the oscillation frequencies used and not on the determination and uncertainties in $\nu_{max}$.

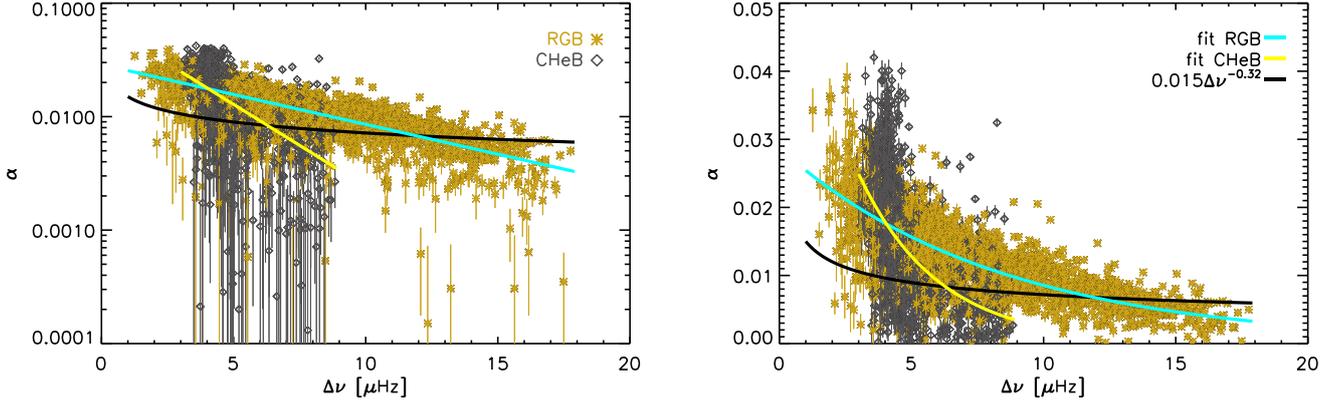

**Fig. 2.** Curvature $\alpha$ as a function of the large frequency separation, with $\alpha$ plotted on a logarithmic scale (left) and linear scale (right). See legend for the meaning of the colours.

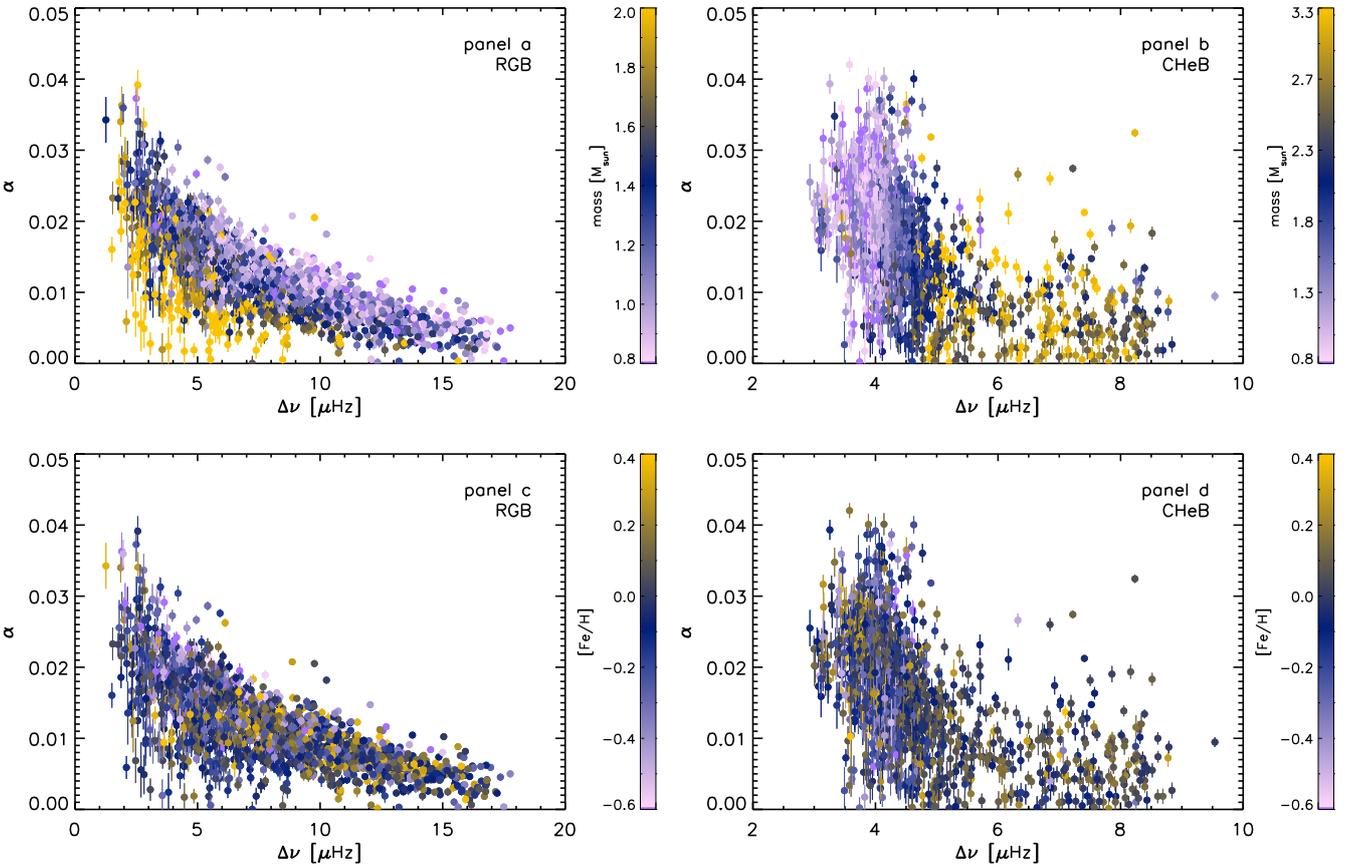

**Fig. 3.** Curvature $\alpha$ as a function of the large frequency separation $\Delta\nu$ for RGB stars (left) and CHeB stars (right), colour-coded by mass (top) and metallicity (bottom).

### 2.2. The $\alpha - \Delta\nu$ relation

In Fig. 2, we show the values of $\alpha$ derived using Eq. (3) as a function of $\Delta\nu$ for red-giant-branch (RGB) stars as well as for stars in the core-helium-burning (CHeB) phase. Based on the morphology of the points in the left panel of Fig. 2, we fit a linear relation in log $\alpha$ as a function of $\Delta\nu$, i.e.,

$$\log \alpha = d_1 \Delta\nu + d_2. \quad (7)$$

In Table 1, we list the values of $d_1$ and $d_2$ for both RGB and CHeB stars.

In Fig. 2, we also show the $\alpha - \Delta\nu$ relationship proposed by Mosser et al. (2012), see Eq. (2). We note that this fit overestimates the values at higher $\Delta\nu$ and underestimates $\alpha$ at lower values of $\Delta\nu$. It should however be remembered that the formulation for $\alpha$ versus $\Delta\nu$ proposed by Mosser et al. (2012) was de-rived over a decade ago at a time when the available timeseries data were of shorter duration than those used in this study. This may have led to a smaller number of radial modes detected and higher uncertainties on the determined frequencies. Indeed, for higher values of $\alpha$ as found at low $\Delta\nu$, a fit to a smaller number of modes is likely to underestimate the value of $\alpha$. It may well be that the six radial modes used in this work are also not sufficient to obtain the actual value of $\alpha$ in these cases. However, this issue remains elusive as the number of modes is limited for evolved stars due to the low radial order. Finally, as mentioned above, we have fit the Universal pattern in a different way, which may impact both $\alpha$ and $\Delta\nu$.

**Table 1.** Parameters of Eq. (7) for RGB and CHeB stars, including the standard deviation of the values around the fit and the number of stars fit.

| evol state | $d_1$ [Ms] | $d_2$ | standard error in $\alpha$ | # of stars |
|---|---|---|---|---|
| RGB | $-0.0526 \pm 0.0002$ | $-1.542 \pm 0.002$ | 0.004 | 2242 |
| CHeB | $-0.1445 \pm 0.0007$ | $-1.170 \pm 0.004$ | 0.006 | 1698 |

### 2.3. Mass and metallicity dependence of $\alpha$

Now that we have shown how $\alpha$ depends on $\Delta\nu$, we turn to other stellar parameters, i.e. mass and metallicity. In Fig. 3, we show the curvature versus large frequency separation colour-coded by mass and metallicity for the RGB and CHeB stars.

Mass dependence: For CHeB stars, it is well known that there is a strong dependence between mass and $\Delta\nu$ (see e.g. Mosser et al. 2014), and indeed we recover this trend in Fig. 3. We do not find any particular variation in $\alpha$ with mass at fixed $\Delta\nu$. For the RGB stars, the situation is different. We find that at any value of $\Delta\nu$, $\alpha$ depends inversely on mass, i.e. $\alpha$ increases with decreasing stellar mass. We investigate this mass dependence using the models later in this paper.

Metallicity dependence: There does not seem to be any dependence of $\alpha$ on metallicity for either the RGB or the CHeB stars. Though, in the case of the CHeB stars, this might be due to observational biases, as this sample is dominated by stars with solar metallicity. Certainly for the higher mass stars in the secondary clump the variation in metallicity is too small to draw any conclusions on a potential metallicity dependence of $\alpha$. For the RGB stars, the spread in metallicity is larger than for the secondary clump, though still relatively narrow with values between $-0.6$ and 0.4 dex. For these values the metallicity indepence seems robust.

## 3. Analysis of curvature in stellar models

To compare the curvature results from the observations with predictions, we also analysed stellar models for the curvature using the Modules for Experiments in Stellar Astrophysics (MESA, Paxton et al. 2011, 2013, 2015, 2018, 2019; Jermyn et al. 2023, and references therein) code version r12778. We used the standard settings, i.e. solar composition from Grevesse & Sauval (1998), mixing-length parameter was kept to the value of two as per the default, convective boundaries according to the Schwarzschild criterion with no overshoot and no diffusion and settling of helium and heavy elements. The oscillation frequencies of radial modes were computed using GYRE (Townsend & Teitler 2013; Goldstein & Townsend 2020, and references therein). We computed solar metallicity tracks with masses 0.8,
1.0, 1.3, 1.6, 1.9 and 2.2 M$_\odot$. Additionally, we computed 1.0 M$_\odot$ tracks with metallicities of [Fe/H] = $-0.5, -0.2, 0.0. 0.2$ and 0.35 dex.

To mimic the observations, we selected the six radial mode frequencies closest to $\nu_{\max}$. We subsequently use the selected frequencies to obtain $\alpha$ from Eq. (3). For the models, we computed $\nu_{\max}$ from the scaling relation, i.e. $\nu_{\max} \propto MR^{-2}T_{\rm eff}^{-0.5}$, with $M$, $R$ and $T_{\rm eff}$ in solar units, where $T_{\rm eff,\odot}$ = 5777 K (Prša et al. 2016), and a reference value for $\nu_{\max}$ of 3100 $\mu$Hz (see Hekker 2020, and references therein for a discussion on the reference values for scaling relations).

To account for uncertainties, we perturbed $\nu_{\max}$ randomly, assuming a 5% uncertainty. We imposed this perturbation to allow a change in the set of six modes that were selected and used to determine the curvature. Additionally, we introduced an uncertainty on the individual frequencies. For this we used the quadratic uncertainty model centred around $\nu_{\max}$ described by Ahlborn et al. (2025). To investigate the effect of these uncertainties on the derived values of $\alpha$ and $\Delta\nu$, we performed a Monte Carlo analysis for some individual models. We found that the resulting uncertainties in $\alpha$ and $\Delta\nu$ align with typical observed uncertainties. See Appendix B for more details. We show the $\alpha$ vs. $\Delta\nu$ results for the stellar models in Fig. 4.

Figure 4 shows that $\alpha$ typically increases with decreasing $\Delta\nu$ for both the RGB and CHeB models. We now also discuss the mass and metallicity trends in these models.

Mass dependence: For the RGB stars, the models show a clear mass trend with lower mass stars having more curvature (larger values of $\alpha$) at the same $\Delta\nu$. To investigate this further, we performed linear fits in $\log\alpha$ versus $\Delta\nu$ for individual mass tracks, see Eq. (7). We did this by fitting both the slope ($d_1$) and the intercept ($d_2$) as well as by fixing the slope and only fitting the intercept. The value to which we fixed the slope was the value of $d_1$ obtained from a fit including all models of all mass tracks. We show the intercepts as a function of mass in Fig. 5. The intercept indeed decreases with mass (this trend is irrespective of the slope), confirming that the curvature decreases with stellar mass. For the CHeB stars the well-known mass dependence with $\Delta\nu$ was recovered (see Fig. 4).

Metalicity dependence: For the stars on the RGB nor for stars in the CHeB phase, is there a trend between the metallicity and $\alpha$.

## 4. Comparing $\alpha$ from observations and models

A comparison of the top rows of Figs 3 and 4 shows that the models qualitatively confirm the trends seen in the observations. To quantify this, we selected two observational samples: a solar-metallicity ([Fe/H] = $0.0 \pm 0.1$ dex) sample, and a solar-mass ($M$ = $1.0 \pm 0.1$ M$_\odot$) sample. We subsequently performed a coarse grid-based modelling in which we selected a model for each observed star in the two samples. For the solar-metallicity sample, we selected models from the [Fe/H] = 0 tracks that are closest to the observed mass and $\Delta\nu$. For the solar-mass sample, we se-lect models from the 1 $M_\odot$ tracks that are closest to the observed metallicity and $\Delta\nu$. In Fig. 6, we show the results for both the observed samples as well as their respective models.

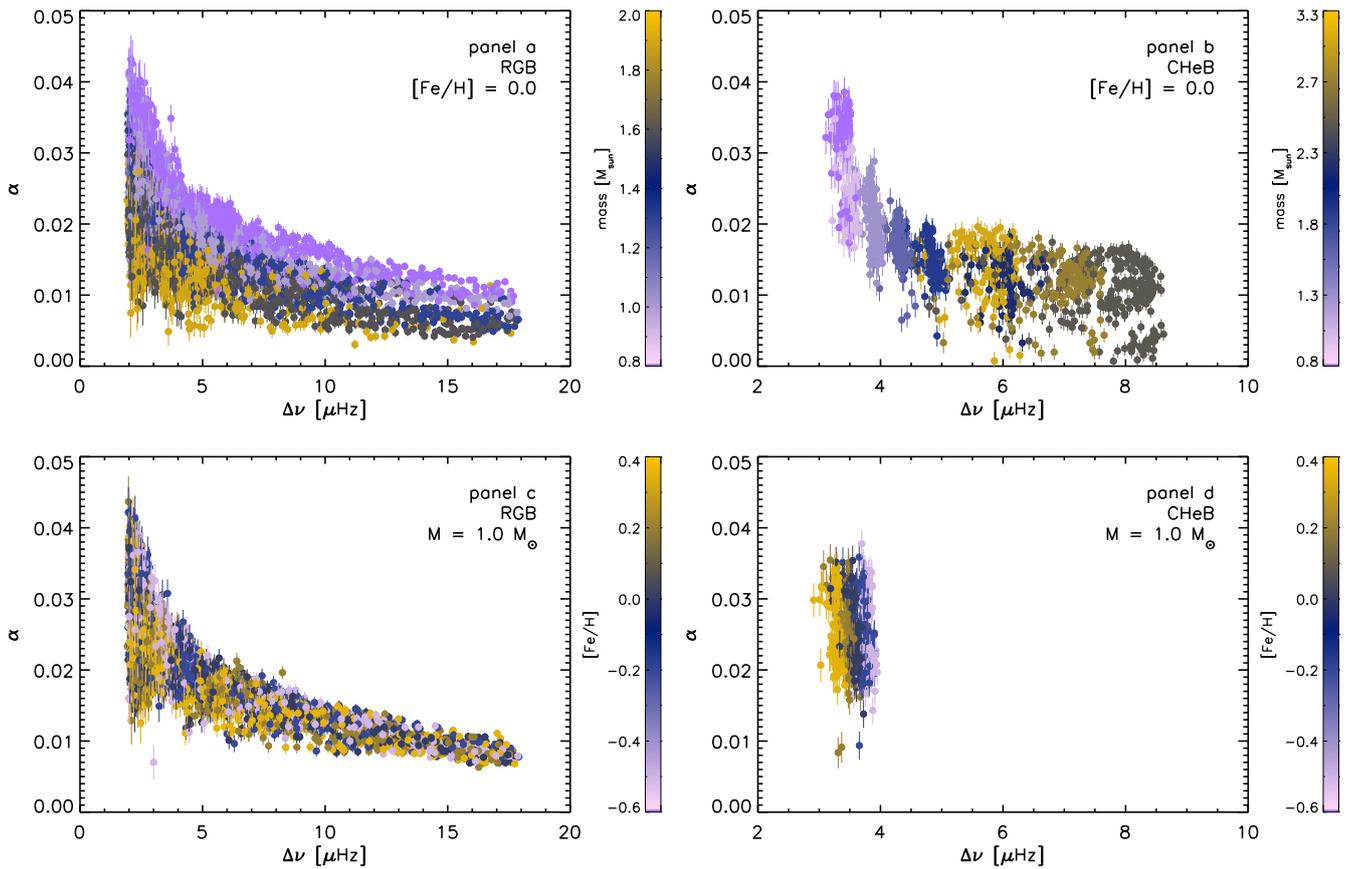

**Fig. 4.** Same as Fig. 3, now for results based on our stellar model computations. Note that in the top panels all models have solar metallicity, while in the bottom panels we show only one mass. The latter is the cause of the reduced frequency range of models in panel d.

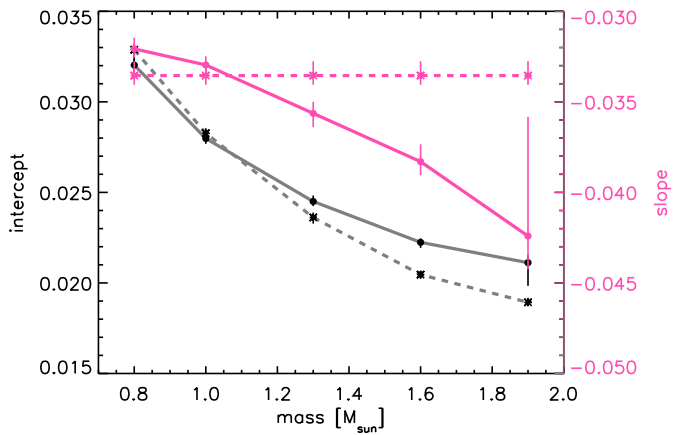

**Fig. 5.** The intercepts (dark gray, left axis) and slopes (pink, right axis) of linear fits in log $\alpha$ vs. $\Delta\nu$ for RGB models as a function of mass. The points show the mean slope and intercept of fits for 1000 perturbations. Where we perturbed both $\nu_{max}$ and the individual frequencies within their uncertainties. The vertical bars indicate the spread in the results of these 1000 perturbations. The points connected with a solid line show the results of fits in which both the slope and the intercept were a free parameter. The dashed lines connect points for fits with a fixed slope. See text for more details and Fig. C.1 for the fits to the different masses.

For the CHeB stars, we find that the fit through the data does indeed overlap with the models, however, the values of $\alpha$ for the solar metallicity models are somewhat higher than those observed. For RGB stars on the other hand, this picture is rather different; the models seem to be offset in $\alpha$ and have a different slope of $\alpha$ vs. $\Delta\nu$. This is most pronounced in Fig. 6d, where the majority of the solar-mass models lie above the fit and the offset from the fit increases with increasing $\Delta\nu$.

It is known that the frequencies of stellar models must be corrected for the surface effect — the inability of the models to properly reproduce the conditions near the stellar surface (Kjeldsen et al. 2008; Ball & Gizon 2014). We therefore investigated if the discrepancy in $\alpha$ between the observations and the models is due to this phenomenon. We did this in two different ways. We first performed a quantitative analysis by including a surface correction $\delta\nu_s$ with an arbitrary shape as a function of $n$ into Eq. (1). From this, one can derive a relation between the modelled value of $\alpha$ ($\alpha_{mod}$) and the corrected $\alpha$ ($\alpha_c$), which should in the ideal case be the same as the observed $\alpha$ ($\alpha_{obs}$):

$$\alpha_{obs} \approx \alpha_c \approx \alpha_{mod}\frac{\Delta\nu_{mod}}{\Delta\nu_{obs}} + \frac{1}{\Delta\nu_{obs}}\frac{d^2\delta\nu_s}{dn^2}, \qquad (8)$$

see Appendix A for the derivation of Eq. (8). The known shape of the surface corrections implies that $\Delta\nu_{mod} > \Delta\nu_{obs}$ (e.g. Ball & Gizon 2014). Thus, for the larger values of $\alpha_{mod}$ compared to $\alpha_{obs}$ to be explained by the surface effect, the second order derivative of $\delta\nu_s$ needs to be negative. We used the Sun, the main-sequence stars from Buchele et al. (2024, 2025) and RGB stars from Ball et al. (2018) to test this (see Appendix A for details).

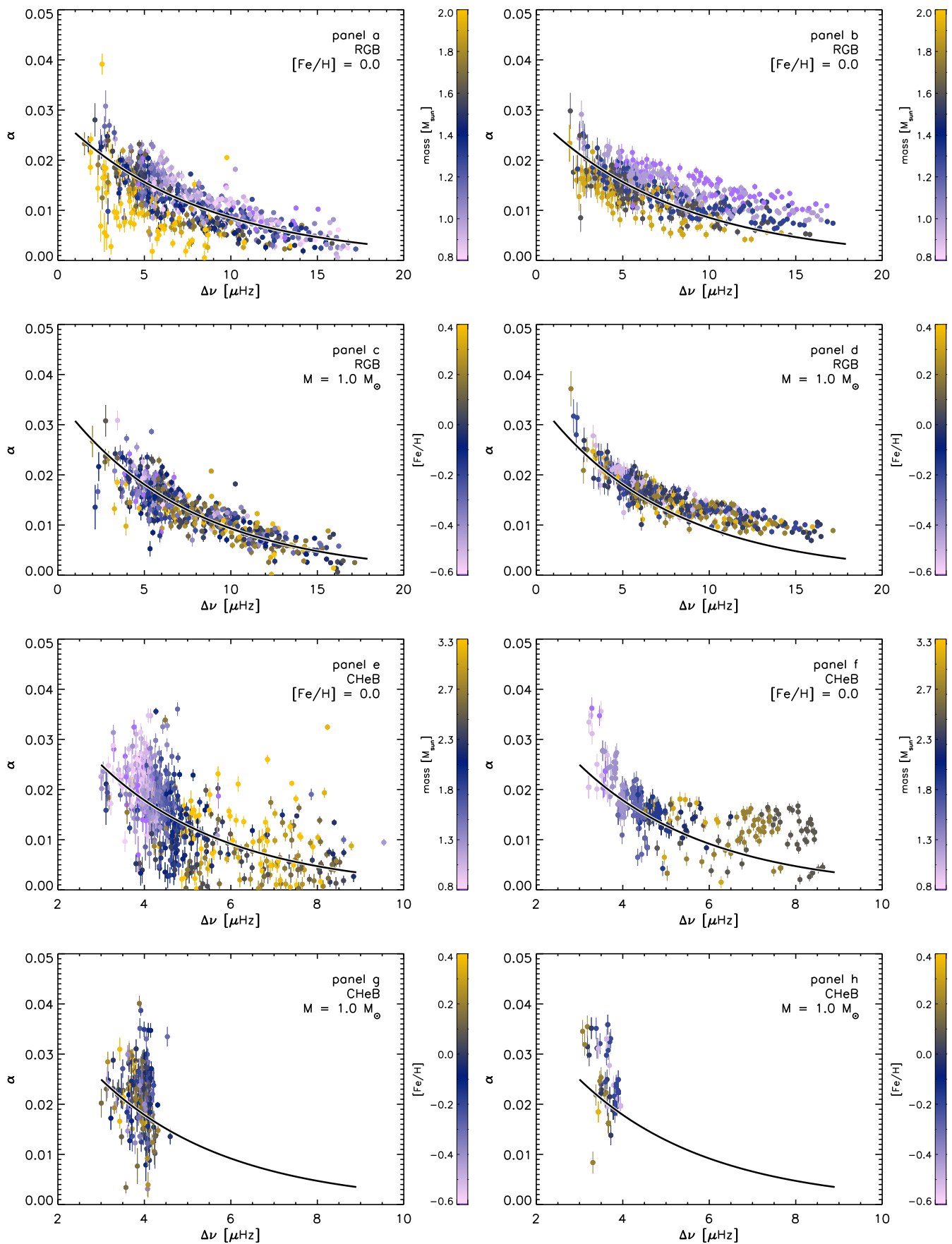

**Fig. 6.** Curvature $\alpha$ as a function of the large frequency separation $\Delta\nu$ for observations (left) and models (right), split for RGB stars (panels a-d) and CHeB stars (panels e-h), colour coded by mass (panels a-b & e-f) and metallicity (panels c-d & g-h) where the colour code is the same for both the observations as the models. The black lines indicate the RGB and CHeB fits from Fig. 2, and Eq. (7) with the coefficients from Table 1. Except for panels c and d where we use a fit through the selected observed values to account for the dependence of $\alpha$ on mass.

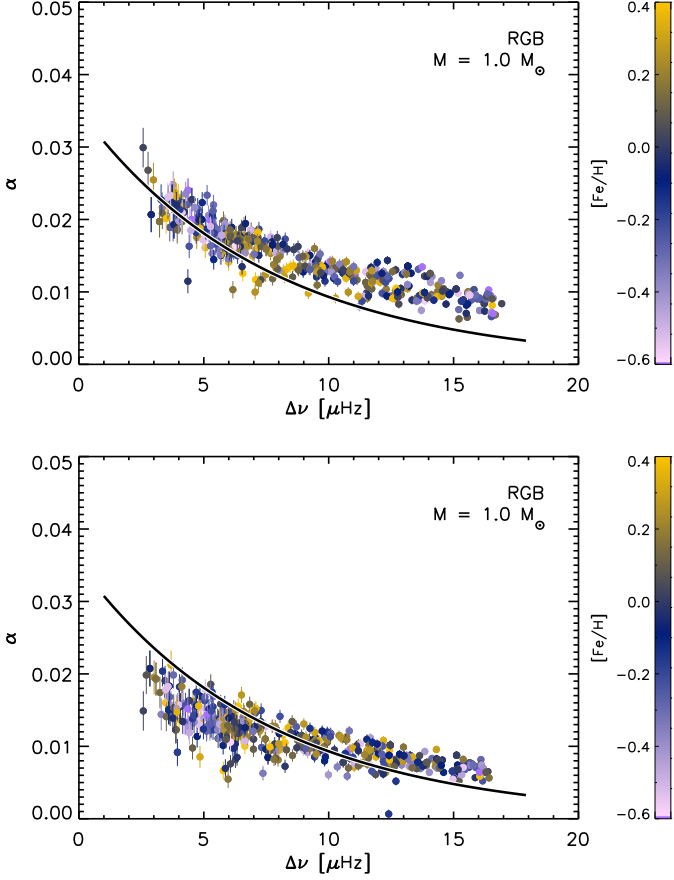

**Fig. 7.** Same as Fig. 6d for uncorrected (top) and surface corrected frequencies (bottom) using models from Li et al. (2023).

Indeed, we find that a negative value for this derivative is likely and thus that the surface effect could be responsible for the discrepancy between $\alpha_{obs}$ and $\alpha_{mod}$. To take this qualitative result one step further and quantify if the surface correction is indeed the culprit, we also used surface corrected models from Li et al. (2023) to investigate the impact of the surface correction on $\alpha$.

Li et al. (2023) performed a study in which they computed the $a_3$ and $a_{-1}$ coefficients for the surface effect described in Ball & Gizon (2014) based on stellar parameters, i.e. surface gravity, effective temperature and metallicity. They also made a grid of models publicly available in which their surface effects are already included. Here we use the models and frequencies from this grid and repeat the analysis of Sect. 3 with these models. We show the $\alpha - \Delta\nu$ relation for both the corrected and uncorrected frequencies as computed by Li et al. (2023) in Fig. 7. The uncorrected results of the top panel of Fig. 7 closely resemble Fig. 6d. For the surface corrected results shown in the bottom panel of Fig. 7 the values of $\alpha$ are indeed reduced. However, the reduction seems to be too strong at lower values of $\Delta\nu$ (i.e. $\Delta\nu \lesssim 7$ $\mu$Hz) and not strong enough for models with $\Delta\nu \gtrsim 12$ $\mu$Hz. This could mean that the surface correction should be altered, or the surface correction is correct and there is some additional physics missing in the models. At the moment we cannot distinguish between these options. However, if it is the latter, it is important to know the part of the star to which $\alpha$ is most sensitive. We discuss this in the next section.

## 5. Dependence of curvature on stellar structure features

To investigate which part of the star the curvature is most sensitive to, we use structure kernels following the approach described by Otí Floranes et al. (2005) and references therein. These authors examined the sensitivities of frequency ratios on the stellar structure. The idea is as follows: frequency differences between two sufficiently similar stellar models (or between an observed star and a sufficiently close model) can typically be expressed in terms of differences in two structure variables, for example squared sound speed ($c^2$) and density ($\rho$). The frequency differences and structure differences are connected through kernel functions. These kernel functions quantify how changes to the stellar structure translate into changes in the pulsation frequencies, i.e. the kernel functions indicate the sensitivity of a structure parameter as a function of some stellar coordinate. The latter is often (though not necessarily) expressed in fractional radius. Using these kernels, differences in frequencies can now be 'translated' into differences in internal structures through their kernels.

To find the sensitivity of $\alpha$, we write Eq. (1) for $\ell = 0$ modes, and define $x = n - n_{max}$ to obtain:

$$\nu_{x+n_{max},0} = \Delta\nu\left(x + n_{max} + \epsilon_p + \frac{\alpha}{2}x^2\right). \quad (9)$$

Taking the derivative with respect to $x$ yields

$$\frac{d\nu_{x+n_{max},0}}{dx} = \Delta\nu + \alpha\Delta\nu\, x \quad (10)$$

where $n_{max}$ and $\epsilon_p$ are assumed to be constant with respect to $x$. To solve this equation, we do need a value for $n_{max}$ which we obtain from $n_{max} = \nu_{max}/\Delta\nu - \epsilon_p$. We checked differences in values for $\Delta\nu$ and $\alpha$ obtained from Eq. (3) and Eq. (10). The difference in $\Delta\nu$ are well below 1%. The differences in $\alpha$ are larger, though at least one order of magnitude smaller than the actual values. Eq. (10) is a linear relationship

$$y = \beta_0 + \beta_1 x \quad (11)$$

with the intercept $\beta_0 = \Delta\nu$ and the slope $\beta_1 = \alpha\Delta\nu$. To obtain $\alpha$ we can divide the slope ($\beta_1$) through the intercept ($\beta_0$):

$$\alpha = \frac{\alpha\Delta\nu}{\Delta\nu} = \frac{\beta_1}{\beta_0}. \quad (12)$$

Following Otí Floranes et al. (2005), we wrote the linearized perturbations in $\alpha$, denoted by $\delta\alpha$, as a function of the perturbations in $\alpha\Delta\nu$ ($\delta(\alpha\Delta\nu)$) and $\Delta\nu$ ($\delta\Delta\nu$) as

$$\delta\alpha = \frac{1}{\Delta\nu}(\delta(\alpha\Delta\nu) - \alpha\delta\Delta\nu). \quad (13)$$

The kernel for differences in $\alpha$, i.e. $\underline{K}_\alpha$, is therefore

$$\underline{K}_\alpha = \frac{1}{\Delta\nu}\left(\underline{K}_{\alpha\Delta\nu} - \alpha\underline{K}_{\Delta\nu}\right) \quad (14)$$

$$= \frac{1}{\Delta\nu}\left(\underline{K}_{\beta_1} - \alpha\underline{K}_{\beta_0}\right), \quad (15)$$

where $\underline{K} \equiv [K^{c^2}, K^\rho]$ with $K^{c^2}$ and $K^\rho$ the sensitivity kernels for sound speed ($c$) and density ($\rho$) respectively (for the derivations of these kernels see e.g. Gough & Thompson 1991; Kosovichev 1999; Basu 2016). These kernels are determined by the structure and mode frequencies of the models[2].

---
[2] In this work, we are interested in only the radial modes, therefore the dependence of mode degree $\ell$ has been omitted.

To obtain $\underline{K}_{\beta_0}$ and $\underline{K}_{\beta_1}$, we used again the linear expression of Eq. (11). This linear regression can conveniently be written in matrix notation:

$$\mathbf{Y} = \beta \cdot \mathbf{X}, \tag{16}$$

with $\mathbf{Y} = (d\nu_1/dx_1, ..., d\nu_N/dx_N)^T$, $\mathbf{X} = (x_1, ...x_N)$ and $\beta = (\beta_0, \beta_1)^T$. The values for $\mathbf{Y}$ were determined as the frequency differences between radial modes with consecutive radial orders and $\mathbf{X}$ are the radial order differences of these eigen frequencies. For this analysis, we selected from the models described in Section 3 a subset with two different masses ($M = 1.0$ and $1.9\ M_\odot$) and a sufficient number of radial modes. That is, we set $N = 6$ to include the six modes closest to $\nu_{\max}$ consistent with what was used for the observations and models presented in Sections 2 & 3 of this work. Following Ahlborn et al. (2022), we determine $\beta$ by means of linear regression and the parameters are estimated as:

$$\beta = (\mathbf{X}^T \cdot \mathbf{X})^{-1}\mathbf{X} \cdot \mathbf{Y} = \mathbf{C} \cdot \mathbf{Y} \tag{17}$$

The matrix $\mathbf{C}$ can be interpreted as a coefficient matrix, i.e. with coefficients $c_{\beta_j}$, determining the impact (weight) of each data point on the derived parameters $\beta_j$, while $\mathbf{Y}$ is the difference in consecutive frequencies. So we can write this as:

$$\beta_j = \sum_{i \in \mathcal{M}} c_{\beta_j,i}(\nu_{i+1} - \nu_i) \tag{18}$$

We can use this to construct sensitivity kernels for the slope $\beta_1$ and the intercept $\beta_0$:

$$\underline{K}_{\beta_j} = \sum_{i \in \mathcal{M}} c_{\beta_j,i} \left(\underline{K}_{i+1}(r) - \underline{K}_i(r)\right), \tag{19}$$

where $\mathcal{M}$ refers to the mode set of interest, and $\underline{K}_i(r)$ are the kernel functions of each oscillation mode $i$.

We computed $\underline{K}_\alpha$ for several models along both the 1 $M_\odot$ and 1.9 $M_\odot$ tracks (see right panel of Fig. 9). For a representative model, we show $\underline{K}_\alpha$ in Fig. 8. The largest amplitude in $\underline{K}_\alpha$ is located relatively close to the stellar surface and overlaps with the dip in $\Gamma_1$ indicating the He I and H I ionisation layer. This is the case for all models we investigated. For reference, we also show the radial eigenfunction of the radial mode closest to $\nu_{\max}$ in Fig. 8. This shows that the period of the eigenfunction at the combined He I and H I ionisation layer is of the same order or larger than the extent of this layer.

## 6. Dependence of $\alpha$ on characteristics of ionisation layers

In this Section, we investigate if we can derive any information about the He I and H I ionisation layers from $\alpha$. To do so, we first return to Eq. 1 and note that this equation is derived by Mosser et al. (2013b) as the observational counterpart of the asymptotic approximation by Houdek & Gough (2007). In Eq. 1 the curvature term replaces the degree independent part of the second order term in Eq. 20 of Houdek & Gough (2007), which depends on the stratification near the surface (Gough 1986, and references therein). Indeed, Fig. 8 confirms the sensitivity of $\alpha$ to the near-surface layers. Additionally, we show in Fig. 8 that the local wavelength of the eigenfunction is of the same order as the extent of the He I and H I ionisation layers. Like for other features that occur on length scales shorter or comparable to the local wavelength, such as the previously studied He II ionisation layer (see e.g. Miglio et al. 2010; Mazumdar et al. 2014;

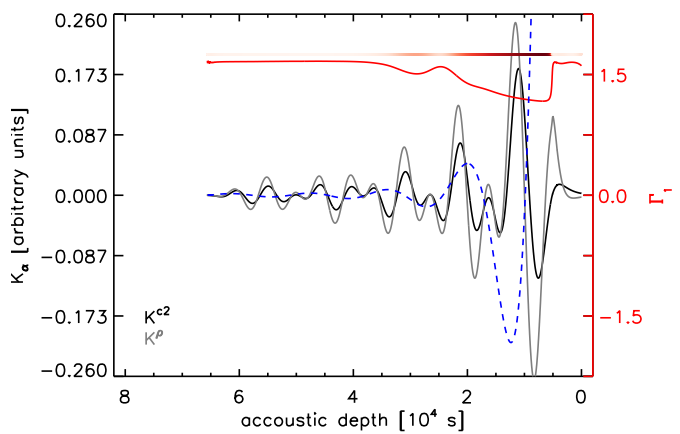

**Fig. 8.** Kernels for $\alpha$ ($\underline{K}_\alpha$) as a function of acoustic depth (black and gray) for a 1 $M_\odot$ model with a $\nu_{\max}$ of about 76 $\mu$Hz. For comparison we also show $\Gamma_1$, the first adiabatic exponent in red, as well as its deviation from 5/3 in the colour-code just above it. For comparison, the blue dashed curve shows the radial eigenfunction of the radial mode closest to $\nu_{\max}$ in arbitrary units.

Broomhall et al. 2014), we can treat the combined He I - H I ionisation layer as a glitch, despite it being a more borderline case. Such a glitch can be described as:

$$\delta\nu = A_g \sin(4\pi\tau_g \nu + \phi), \tag{20}$$

with $A_g$ the amplitude, $\tau_g$ the period and $\phi$ the phase, where the amplitude would to some extent correlate with the abundance of the hydrogen and helium in the ionisation zone (see Broomhall et al. 2014, for an analysis of helium abundances from the amplitudes of He II glitch signatures), and $\tau_g$ would correlate with the acoustic depth of these ionisation layers. We note here, that the He II ionisation layer and base of the convection zone are located deeper in the star than the He I and H I layers. At these larger depths, $\alpha$ has much less sensitivity, and these deeper features appear as variations with shorter periods (see e.g. the deviations of the red points from the red dashed line in Fig. 1), which can be neglected when studying $\alpha$. To explore this further, we replace the curvature term in Eq. (1) by Eq. (20) and obtain for radial modes

$$\nu_{n,0} = \Delta\nu(n + \epsilon_p) + A_g \sin(4\pi\tau_g \nu + \phi). \tag{21}$$

To derive values for $A_g$ and $\tau_g$, we performed a Monte Carlo fit with 5 000 iterations to perturbed frequencies of the same models as used in Section 5. We present the results of this fitting in Fig. 9, where we compare these values with the ones obtained through the methods described in Sections 2 & 3. From this, we find indeed that the values for $\tau_g$ are in line with the region close to the surface as identified by the kernel sensitivity, i.e. between $0.1 \times 10^4$ s and $2 \times 10^4$ s as also shown in Fig. 8. Furthermore, the amplitudes $A_g$ decrease with mass while $\tau_g$ increases with mass, at a similar value for $\alpha$ (see Fig. 9).

To quantify this further we attempted a Taylor expansion of the sinusoid up to the quadratic term to mimic Eq. (3), in Eq. (21), for which we converted the sinusoid to a cosine[3] and use $-A_g \cos x = -A_g(1 - x^2/2)$, with $x = 4\pi\tau_g \nu + \phi$ and $\nu = \Delta\nu(n + \epsilon_p)$ to obtain:

$$\nu_{n,0} = A_T n^2 + B_T n + C_T \tag{22}$$

---
[3] The $\pi/2$ factor to convert the sin of Eq. (20) to cos has been absorbed in the phase $\phi$.

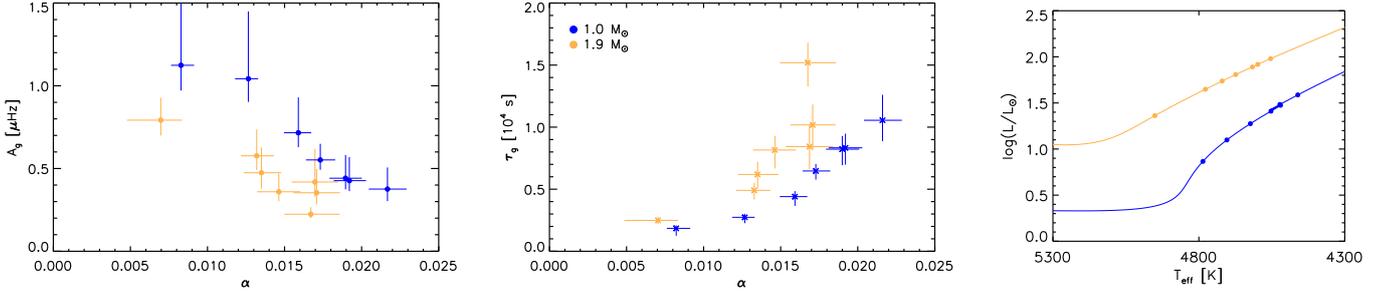

**Fig. 9.** The amplitude ($A_g$, left) and acoustic depth ($\tau_g$, centre) as a function of $\alpha$ for direct fits of Eq. (20) to the frequencies of models of 1 M$_\odot$ (blue) and 1.9 M$_\odot$ (orange). The bars indicate the 25-75 percentile of the distribution of the results of 5 000 perturbations. The tracks evolve from left to right. The positions of the models in the Hertzsprung-Russell diagram are indicated with the dots in the right panel. The lines are the full tracks shown to guide the eye.

with

$$A_T = \frac{A_g}{2}\left(4\pi\tau_g\Delta\nu\right)^2 \quad (23)$$

$$B_T = \Delta\nu + \frac{A_g}{2}\left(8\pi\tau_g\Delta\nu(4\pi\tau_g\Delta\nu\epsilon_p + \phi)\right) \quad (24)$$

$$C_T = \Delta\nu\epsilon_p - A_g + \frac{A_g}{2}\left(4\pi\tau_g\epsilon_p\Delta\nu(4\pi\tau_g\epsilon_p\Delta\nu + 2\phi) + \phi^2\right) \quad (25)$$

Eq. (22) has the same shape as Eq. (3). As we derived $\alpha$ from the quadratic term in Eq. (3) only, we now also only included the quadratic term. This provides:

$$\frac{A_g}{2}\left(4\pi\tau_g\Delta\nu\right)^2 = \frac{\alpha}{2}\Delta\nu \quad (26)$$

thus

$$\alpha = A_g(4\pi\tau_g)^2\,\Delta\nu \quad (27)$$

as indeed both $A_g$ and $\tau_g$ are included, this emphasizes that $\alpha$ is determined by both the period and the amplitude of the He I and H I ionisation layers[4]. Upon checking this equality for the models, i.e $\alpha$ as determined in Section 3 and from using the right-hand side of Eq. (27), we indeed find reasonable agreement, see Fig. 10, empirically justifying the glitch approach. For higher values of $\alpha$ ($\alpha > 0.02$) the agreement is not as good, potentially as for these higher values of $\alpha$ the six modes may not be enough to quantify the strong curvature and/or the glitch analysis with a Taylor expansion with only the quadratic term may no longer provide a proper approximation.

These results, in which both the amplitude and the depth of the ionisation zones are intimately connected, prevent us from a direct determination of the location or the abundance of the near surface ionisation zones from $\alpha$. However, if this degeneracy can be lifted and either the abundance or the location of the near-surface ionisation layers can be determined by other means, $\alpha$ can aid in finding the remaining unknown parameter.

## 7. Discussion and conclusions

In this work, we have presented a way to obtain the curvature of the radial modes i.e., the quadratic deviation of the modes from uniform spacing in a homogeneous way. We did so without the need to compute the radial order of the frequency of maximum oscillation power ($n_{max}$) explicitly. The advantage of this approach is that the result relies only on the oscillation modes included. The observed values of the curvature have lower values than the ones obtained from stellar models. We investigated whether this is a result of the so-called surface effect. Indeed if we include the surface effect, the overestimation of the models reduces. However, a frequency-dependent offset remains. This may be showing that the surface correction may need to be improved, or it may be due to other physical shortcomings in the models. To investigate this further, we computed the sensitivity of the curvature, which is dominant in the combined He I and H I ionisation zone close to the stellar surface. The results of the sinosoidal approximation indeed confirm this location. This proximity to the surface most likely means that this is entangled with the surface effect, although we clearly showed that a state-of-the-art treatment of the surface effect does not fully mitigate the discrepancy between the observations and the models. From a Taylor expansion, we find that the curvature depends on both the amplitude and the period of the sinusoid. Hence, both the location and abundance of H and He play a role in the curvature.

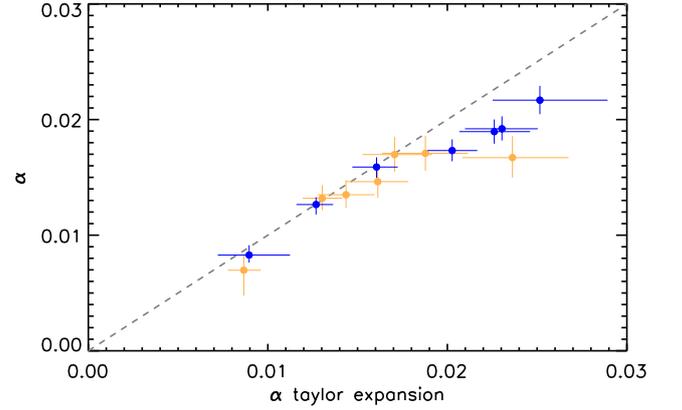

**Fig. 10.** A comparison of $\alpha$ determined via Eq. (3) vs. the right-hand side of Eq. (27). The colours and lines have the same meaning as in Fig. 9.

*Acknowledgements.* We thank the referee and editor for their comments and work, which improved the manuscript significantly. We acknowledge funding from the ERC Consolidator Grant 'DipolarSound' (grant agreement # 101000296). SB is partially supported by NSF grant AST-2205026. YE acknowledges funding from the Science and Technology Facilities Council (STFC).

---

[4] We note here, that when applying this to observations, the surface effect will have to be accounted for.

## Appendix A: Surface effect

To gain insight in the direction and amount of the surface effect on $\alpha$, we look at the asymptotic description of radial modes (see Eq. (1)) and compute its derivatives to $n$:

$$\nu_n = \Delta\nu(n + \epsilon_p + \frac{\alpha}{2}(n - n_{\max})^2), \quad (A.1)$$

$$\frac{d\nu_n}{dn} = \Delta\nu(1 + \alpha(n - n_{\max})), \quad (A.2)$$

$$\frac{d^2\nu_n}{dn^2} = \alpha\Delta\nu. \quad (A.3)$$

We subsequently include a surface correction $\delta\nu_s$, which is in this reasoning an arbitrary function of $n$ which remains small enough such that the asymptotic description of radial modes is still valid. We call the corrected curvature, which should resemble the observations, $\alpha_c$. From this, it follows that:

$$\nu_n + \delta\nu_s = \Delta\nu_c(n + \epsilon_{p,c} + \frac{\alpha_c}{2}(n - n_{\max})^2), \quad (A.4)$$

$$\frac{d(\nu_n + \delta\nu_s)}{dn} = \Delta\nu_c(1 + \alpha_c(n - n_{\max})), \quad (A.5)$$

$$\frac{d^2(\nu_n + \delta\nu_s)}{dn^2} = \alpha_c\Delta\nu_c. \quad (A.6)$$

Now splitting the first term, we find

$$\frac{d^2\nu_n}{dn^2} + \frac{d^2\delta\nu_s}{dn^2} = \alpha_c\Delta\nu_c,$$

$$\alpha\Delta\nu + \frac{d^2\delta\nu_s}{dn^2} = \alpha_c\Delta\nu_c. \quad (A.7)$$

Thus,

$$\alpha_c = \alpha\frac{\Delta\nu}{\Delta\nu_c} + \frac{1}{\Delta\nu_c}\frac{d^2\delta\nu_s}{dn^2}. \quad (A.8)$$

We investigated the value of $d^2\delta\nu_s/dn^2$ from observational results. For this, we used solar data from BISON (Basu et al. 2009) and model S (Christensen-Dalsgaard et al. 1996), as well as data and models of main-sequence stars from (Buchele et al. 2024, 2025, and references therein), and red-giant stars with their models from Ball et al. (2018). For the Sun we computed the surface effect, using the formalism by Ball & Gizon (2014) and use 11 radial orders ($n = [18, 28]$) to compute $\alpha$, $\Delta\nu$ and the median value of $d^2\delta\nu_s/dn^2$ across the different frequencies. For both other datasets, we used the surface corrected and uncorrected frequencies as per the publications. We did not adhere to the six radial modes, as we have done for the red-giant stars we analysed in this work for two reasons: 1) we are not interested in $\alpha$, but in the second order derivative of the correction and 2) most of the fitted stars are main-sequence stars where more radial modes were observed and may also be necessary for a proper determination of $\alpha$. Figure A.1 shows the results we obtain for the median value of $d^2\delta\nu_s/dn^2$. Indeed, a negative value of $d^2\delta\nu_s/dn^2$ is most likely, and becomes more likely with increasing $\Delta\nu$. This negative value could counteract the effect of the larger value of $\Delta\nu$ in uncorrected models and lead to a smaller value of the corrected $\alpha$ compared to the uncorrected $\alpha$.

## Appendix B: Monte Carlo test for individual models

Here, we investigated the spread obtained in $\Delta\nu$ and $\alpha$ following the 5% uncertainty we induced on $\nu_{\max}$ and the uncertainties on the individual frequencies as per Ahlborn et al. (2025). For this investigation, we take three models with different values of $\nu_{\max}$, see Fig. B.1. The induced spread on $\nu_{\max}$ and the fact that the oscillations are chosen closest to $\nu_{\max}$, means that we select different sets of frequencies. This results in different values of $\Delta\nu$, each with narrow distributions with minimal or no overlap. Hence, the values are precise to roughly <1%, consistent with observations, though valid for a specific set of frequencies. The values of $\alpha$ are also affected, although the different distributions tend to overlap.

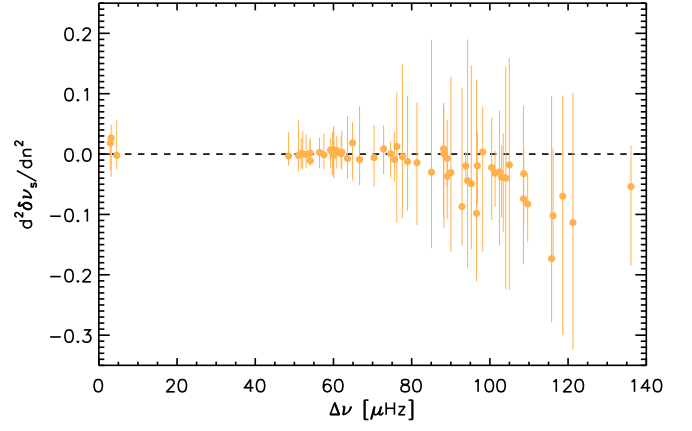

**Fig. A.1.** The value of $d^2\delta\nu_s/dn^2$ per star as a function of $\Delta\nu$. The solid dots indicate the median value, and the vertical lines the spread of $d^2\delta\nu_s/dn^2$ as a function of frequency.

## Appendix C: Mass dependence of $\alpha$

Here, we provide the $\alpha$ vs. $\Delta\nu$ values of models with the fits for different masses. The slopes and intercepts of the fits shown in Fig. C.1, have been summarised in Fig. 5.

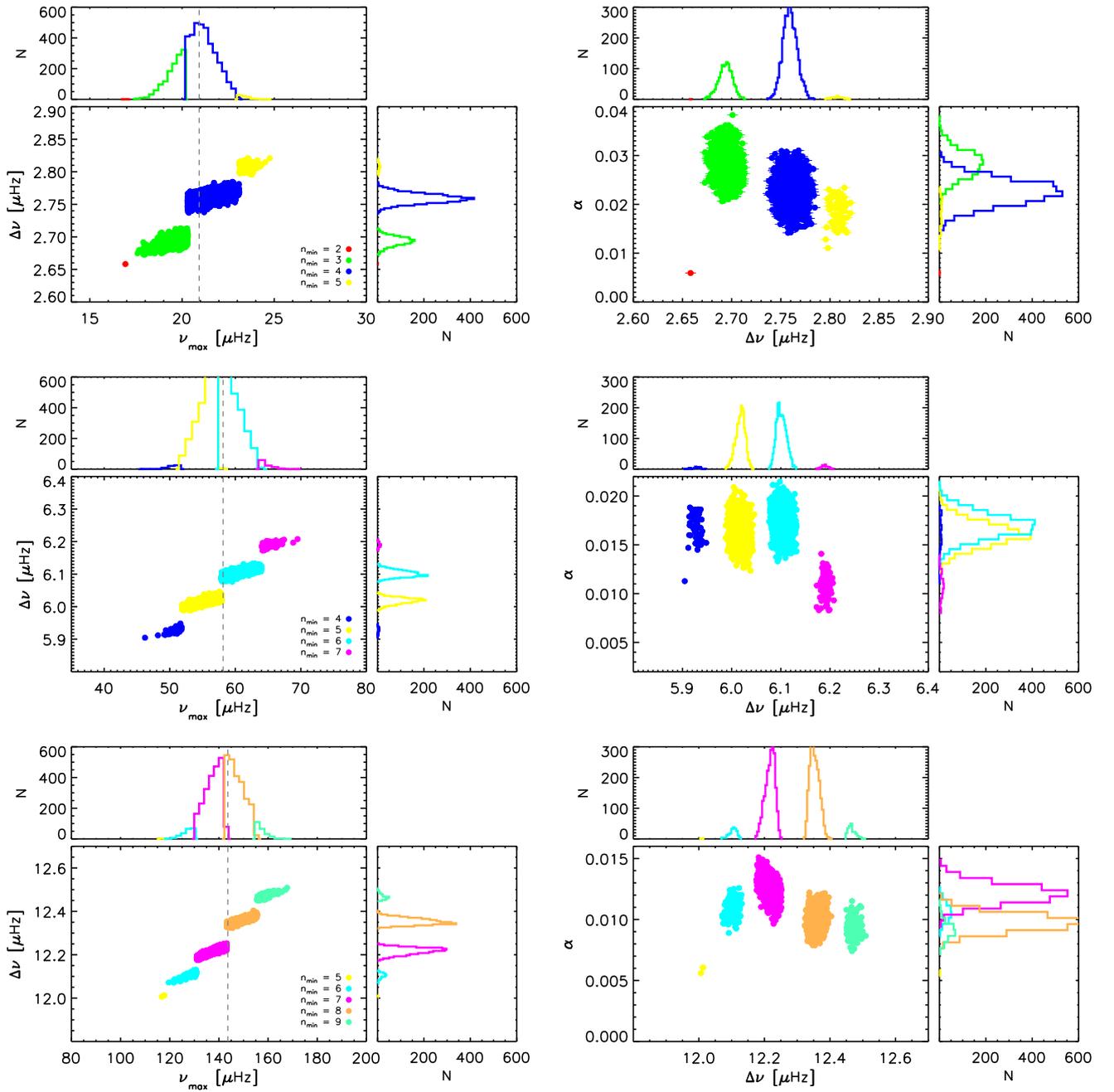

**Fig. B.1.** For three models of one solar mass and solar metallicity the results of Δν versus $\nu_{max}$ (left) and α vesus Δν (right) for 5000 perturbations of 5% uncertainty in $\nu_{max}$ and the individual frequencies. The different colours indicate different sets of modes (see legend where the *n* value of the lowest frequency mode of the set is indicated) that are selected to be the ones closest to the perturbed $\nu_{max}$ value. The vertical gray dashed line indicates the nominal value of $\nu_{max}$ of the respective model. Histograms of the parameters are shown in the top and right panels, respectively, with the same colour-code as the main panels.

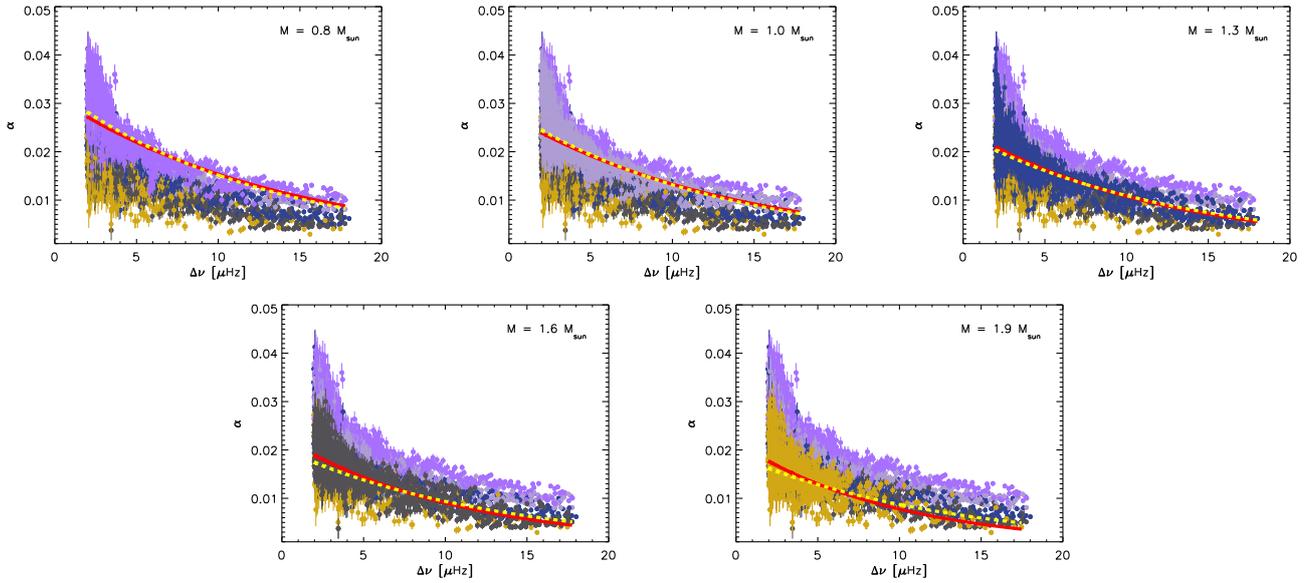

**Fig. C.1.** Curvature $\alpha$ as a function of the large frequency separation $\Delta\nu$ for RGB stars of solar metallicity, colour-coded by mass. In each panel, the mass that is indicated in the legend is shown on top of all other masses. The red solid and yellow dashed lines are fits to all models of that mass. For the solid line both the slope and the intercept were free parameters, while for the dashed line the slope was kept fixed to the slope of the fit to all models.